\documentclass[global,twocolumn]{svjour}
\usepackage{epsfig}
\usepackage{bbm,amsmath,amssymb}
\usepackage{graphics}
\journalname{Applied Physics B}

\newcommand{\bra}[1]{\ensuremath{\left\langle{#1}\right\vert}}

\newcommand{\ket}[1]{\ensuremath{\left|{#1}\right\rangle}}

\def\be{\begin{equation}}
\def\ee{\end{equation}}
\def\eea{\end{eqnarray}}
\def\bea{\begin{eqnarray}}

\newcommand{\ex}[1]{\ensuremath{\left\langle{#1}\right\rangle}}

\newcommand{\exs}[1]{\ensuremath{\langle{#1}\rangle}}
\newcommand{\eins}{\ensuremath{\mathbbm 1}}
\newcommand{\vr}{\ensuremath{\varrho}}

\newcommand{\ketbra}[1]{\ensuremath{| #1 \rangle \langle #1 |}}

\newcommand{\kommentar}[1]{}

\newcommand{\openone}{\ensuremath{\mathbbm 1}}

\begin{document}
\title{Separability criteria and entanglement witnesses for symmetric quantum states}
\author{G\'eza T\'oth\inst{1,2,3} \and Otfried G\"uhne \inst{4,5}
}                     
%
%
\institute{Department of Theoretical Physics,
The University of the Basque Country, P.O. Box 644,
E-48080 Bilbao, Spain\and
IKERBASQUE, Basque Foundation for Science, E-48011 Bilbao, Spain\and
Research Institute for Solid State Physics and Optics,
Hungarian Academy of Sciences,
P.O. Box 49, H-1525 Budapest, Hungary\and
Institut f\"ur Quantenoptik und Quanteninformation,
\"Osterreichische Akademie der Wissenschaften,
A-6020 Innsbruck, Austria\and
Institut f\"ur Theoretische Physik, Universit\"at Innsbruck,
Technikerstra{\ss}e 25, A-6020 Innsbruck, Austria}

%
%

\maketitle

\begin{abstract}
We study the separability of symmetric bipartite quantum
states and show that a single correlation measurement is
sufficient to detect the entanglement of any bipartite
symmetric state with a non-positive partial transpose.
We also discuss entanglement conditions and entanglement
witnesses for states with a positive partial transpose.
\end{abstract}

\section{Introduction}
\label{intro}

Entanglement is a valuable resource for quantum information
processing and its characterization is one of the central problems
in this field. Despite an enormous progress in the last years
the separability problem, i.e., the question whether a given
quantum state is entangled or separable,
remains a major challenge
\cite{HH09,GT09}.

One of the most efficient entanglement condition is the
one based on the positivity of the partial transpose (PPT) of
the density matrix \cite{AP96,AP96b}. If the partial transpose has a negative eigenvalue then
the quantum state is entangled. The PPT condition is a necessary and sufficient condition
for small systems (qubit-qubit and qubit-qutrit),
while for larger systems it does not detect all entangled states.
That is, there are entangled states that have a positive partial transpose \cite{bound,boundb}.
They are called bound entangled since no entanglement can be distilled
from them by local operations and classical communication, even if
several copies of the state are available. There is an extensive literature
on how to verify that a PPT state is indeed
entangled. Clearly, the most efficient entanglement criterion,
the PPT criterion,
cannot be used for that, and other entanglement conditions must be applied.
One of the most often used ones is the Computable Cross Norm-Realignment
(CCNR) criterion \cite{ccnr,cnnrb}. Covariance matrix criteria \cite{cmc,cmcb} can also be used.
They are known to be more powerful
than the CCNR criterion. Moreover, there are
other conditions that work only for symmetric systems, and they
appear as possible candidates for the detection of PPT entanglement,
since their definition seems to be independent from the PPT condition \cite{UU07,UU07b}.

Here we consider entanglement detection in symmetric systems.
In Ref.~\cite{TG09}, it has already been shown that the entanglement conditions mentioned above
coincide for bipartite states of the symmetric subspace. In this paper, we present
an alternative proof of this fact that is based on simple matrix manipulations,
without a reference to a more complex theory of various entanglement conditions.
We will also examine how this equivalence is reflected in the Schmidt decomposition
of symmetric states. This finding might be useful in future for studying bound
entangled states. Finally, we will show how to construct entanglement witnesses for
bipartite symmetric bound entangled states.

\section{Equivalence of several entanglement criteria}

Let us first clarify what is meant by symmetry and symmetric states.
A bipartite quantum state $\vr$ is {\it symmetric} if $\vr=F\vr=\vr F,$
where $F$ is the flip operator exchanging the two qudits. This is the
state of two bosonic particles. On the other hand, a state $\vr$ is called
{\it permutationally invariant} if $F\vr F=\vr.$ Symmetric states are a
subset of permutationally invariant states. Most of this paper is concerned
with symmetric states.

In general, one can write the density matrix for the state of two $d$-state
systems as
\begin{equation}
\label{rho1}
{\vr}=\sum_{k,l,m,n} \vr_{kl,mn} \ket{k}\bra{l}\otimes\ket{m}\bra{n}.
\end{equation}
For such a density matrix, the partially transposed matrix (with respect to Alice's system)
and the realigned matrix are defined as
\begin{equation}
\label{rhoT1}
    {\vr}^{T_A}=\sum_{k,l,m,n} \vr_{lk,mn} \ket{k}\bra{l}\otimes\ket{m}\bra{n},
\end{equation}
and
\begin{equation}\label{rhoR}
    {\vr}^{R}=\sum_{k,l,m,n} \vr_{km,ln} \ket{k}\bra{l}\otimes\ket{m}\bra{n},
\end{equation}
respectively.
For separable states (that is, states that can be written as
$\vr=\sum_k p_k \ketbra{\alpha_k}\otimes\ketbra{ \beta_k}$
with some probability distribution $\{p_k\}$), the criterion of the positivity of the partial
transpose (PPT) states that ${\vr}^{T_A}$ has no negative eigenvalues \cite{AP96}, while the computable
cross norm or realignment (CCNR) criterion states that the trace norm of
${\vr}^{R}$ is not larger than one  \cite{ccnr}.

Moreover, we will use later the expectation value
matrix \footnote{For similar constructions see \cite{evmatrix,evmatrixb}.} given as
\begin{equation}
\label{eta}
{\eta}(\vr, \{ Q_{kl} \})=\sum_{k,l,m,n} {\rm Tr}(\vr Q_{kl} \otimes Q_{mn}^\dagger) \ket{k}\bra{m}\otimes\ket{l}\bra{n},
\end{equation}
where $Q_{kl}$ are $d^2$ pairwise orthogonal matrices defined as
\begin{equation}\label{locorthog}
    Q_{kl}=\ket{k}\bra{l}.
\end{equation}
$Q_{kl}$ provide a full basis for constructing operators on a $d$-state
system. Pairwise orthogonality means ${\rm Tr}(Q_{kl}^\dagger Q_{mn})=\delta_{km}\delta_{ln}.$
Note that not all $Q_{kl}$ are hermitian. Note also that in Eq. (\ref{eta}) the indices are
chosen such that $k,l$ can be viewed as a common row index and $m,n$ act as a column
index.

After these general definitions, we use the bosonic symmetry of
the quantum state and can formulate:

{\bf Observation 1.}
For states in the symmetric subspace, we have
\begin{equation}
\vr_{kl,mn}=\vr_{ml,kn}=\vr_{kn,ml}=\vr_{mn,kl}.
\end{equation}
Hence, we obtain that
\begin{equation}\label{rhoR2}
    F{\vr}^{R}={\vr}^{T_A}={\eta}(\vr,\{Q_{kl}\}).
\end{equation}
where $F$ is the flip operator. Since $F$ is unitary, the singular
values of ${\vr}^{R}$ equal the absolute values of the eigenvalues of ${\vr}^{T_A}$ and
${\eta}(\vr).$ Especially, since ${\rm Tr}({\vr}^{T_A})=1,$ a negative
eigenvalue of ${\vr}^{T_A}$ implies that the trace norms of ${\vr}^{T_A}$
and ${\vr}^{R}$ are larger than one and vice versa.

The eigenvalues of ${\eta}(\vr)$ do not depend on the particular choice of the
full set of local orthogonal matrices $Q_{kl}.$ We can take any other full set of
basis matrices, obtained from $Q_{kl}$ as
\begin{equation}
M_{kl}=\sum_{m,n} U_{kl,mn} Q_{mn},
\end{equation}
where $U$ is a unitary matrix. The eigenvalues of ${\eta}(\vr,\{M_{kl}\})$ are the same
as that of ${\eta}(\vr,\{Q_{kl}\})$ since ${\eta}(\{M_{kl}\}) = U {\eta}(\{Q_{kl}\}) U^\dagger.$
In particular, it is reasonable to choose $M_{kl}$ to be hermitian. Then, they correspond to
local orthogonal observables \cite{locortog}. Thus, the elements of ${\eta}(\vr, \{M_{kl}\})$
come from correlation measurements, that is, mean values like $\exs{A \otimes B}$ and we
can formulate:

{\bf Observation 2.}
Based on Observation 1, we can say that for every symmetric state $\vr$
\begin{equation}
\label{eig}
\Lambda_{\min}({\vr}^{T_A})  \le  \exs{A\otimes A}_{\vr},
\end{equation}
where $\Lambda_{\min}(X)$ denotes the smallest eigenvalue of $X,$
$A$ is an observable,
and ${\rm Tr}(A^2)=1.$ For an appropriately
chosen $A,$ equality holds in Eq.~(\ref{eig}).
This makes it possible
to detect all states with a non-positive partial transpose as
entangled with a single correlation measurement, provided
one knows for sure  that the state is  symmetric.

{\bf Observation 3.}
If a symmetric quantum state $\vr$ is PPT \cite{AP96}, then
$\exs{A^T \otimes A}\ge 0$ for all observables $A$.
This can be seen noting that $\vr^{T_A}$ is also a valid density matrix
with the PPT property, and Observation 2 applies for it.

We can also define $\eta(\vr,\{R_k\})$ for observables $R_k$ that are not
pairwise orthogonal. If $\eta(\vr,\{R_k\})\ngeqslant 0$ then there is an $A$
such that $\exs{A\otimes A} < 0$ and thus
$\eta(\vr,\{M_k\})\ngeqslant 0.$
On the other hand, if the $R_k$ operators are sufficient to construct any observable,
then the converse is also true and $\eta(\vr,\{M_k\})\ngeqslant 0$ implies
$\eta(\vr,\{R_k\})\ngeqslant 0.$

{\bf Observation 4.}
Let us define the correlation matrix $C$ via its entries
as
\be
C_{kl}(\vr)=\exs{M_k \otimes M_l}-\exs{M_k\otimes \eins}\exs{\eins \otimes M_l}.
\ee
For symmetric states
\be
C(\vr,\{M_k\})\ge 0
\iff
\eta(\vr,\{M_k\}) \ge 0
\iff
\vr^{T_A}\ge 0.
\ee
The condition $C(\vr,\{M_k\})\ngeqslant 0$ has been presented as
an entanglement condition in Ref.~\cite{UU07}. One would expect
that this condition is stronger than $\eta(\vr,\{M_k\})\ngeqslant 0.$
However, $C(\vr,\{M_k\})=\eta(\vr,\{M_k-\exs{M_k}\cdot \openone\}).$
Thus, $\eta(\vr,\{M_k\}) \ngeqslant 0$ for a tomographically complete
set $\{M_k\}$ implies $C(\vr,\{M_k\})\ngeqslant 0.$  Note that
$\{M_k-\exs{M_k}\cdot \openone\}$ are not pairwise orthogonal
observables. The rest of Observation 4 is a direct consequence of
Observation 1.

{\bf Observation 5.}
For symmetric states, we have further that
\be
\Vert C(\{M_k\}) \Vert_1^2 \leq [1-{\rm Tr}(\vr_A^2)][1-{\rm Tr}(\vr_B^2)]
\iff
\eta(\{M_k\}) \ge 0.
\ee
For general states, however, the first condition is a separability
criterion that follows from the so-called covariance matrix criterion.
It is stronger than the CCNR criterion and independent of the
PPT criterion \cite{cmc}.
To see this, note that if $C \geq 0$ then we can use
that $\Vert C(\{M_k\}) \Vert_1 = {\rm Tr}(C) =1-{\rm Tr}(\vr_A^2)$
\cite{TG09}.
The other direction follows from the fact that the first condition is
stronger than the CCNR criterion.

In summary, based on Observations 1-5, the following separability
conditions are equivalent for symmetric states:
\begin{itemize}
\item[(i)]  the PPT condition, that is, $\vr^{T_A}\ge 0$ \cite{AP96},
\item[(ii)] the CCNR criterion given as $\vert\vert \rho^R \vert\vert_1 \le 1,$
where $\vert\vert.\vert\vert_1$ is the trace norm \cite{ccnr},
\item[(ii)]  $\exs{A\otimes A}\ge 0$ for every observable $A,$
\item[(iv)] $\eta(\vr,\{M_k\})\ge 0,$
\item[(v)] $C(\vr,\{M_k\})\ge 0,$
\item[(vi)] $\Vert C(\{M_k\}) \Vert_1^2 \leq [1-{\rm Tr}(\vr_A^2)][1-{\rm Tr}(\vr_B^2)].$
\end{itemize}
If any of these conditions is violated then all the others are also violated, and state $\vr$
is entangled.

\section{Schmidt decomposition}

The observations of the previous section are connected to the Schmidt decomposition of
symmetric states. As shown in Ref.~\cite{TG09}, symmetric states can always be decomposed as
\begin{equation}
    \vr=\sum_k \Lambda_k M_k \otimes M_k, \label{Schmidt}
\end{equation}
where $M_k$ are local orthogonal observables ${\rm Tr}(M_kM_l)=\delta_{kl}.$
This is almost a Schmidt decomposition, with the exception that $\Lambda_k$
can also be negative. Since it will make our discussion easier, we will refer to
Eq.~(\ref{Schmidt}) as the Schmidt decomposition and do not absorb the sign
of $\Lambda_k$ in one of the $M_k$ matrices.

The flip operator can be written as $F=\sum_k M_k \otimes M_k$ \cite{TG09}. For symmetric
states $\ex{F}=1,$ hence $\sum_k \Lambda_k=1.$ From Observation 1, we know then that
$\vr$ is PPT if and only if $\ex{A\otimes A}\ge 0$ for all $A$ observables. Hence,
it is easy to see that  $\vr$ is PPT if and only if all $\Lambda_k$ are nonnegative.
Interestingly, in this case $\Lambda_k$ satisfy the conditions for being a probability
distribution, and Eq.~(\ref{Schmidt}) presents a certain kind of quasi-mixture.
In another context, similar relations between quasi-probability distributions and separability have appeared
before in the literature \cite{sanpera,sanperab,sanperac}.

Besides states with a bosonic symmetry, Ref.~\cite{TG09} was also concerned with
permutationally invariant states. It has been shown that if a state is permutationally
invariant, then it can still be decomposed as in Eq.~(\ref{Schmidt}), and
$-1 \le \sum_k \Lambda_k \le 1.$ For such states, it has also been shown that if
all $\Lambda_k$ are nonnegative (or, equivalently, $\ex{A\otimes A}\ge 0$ for all $A$)
then the state does not violate the CCNR criterion. However, the converse is not true.

{\bf Observation 6.}
Expanding the argument, let us consider an operator given as
\begin{equation}
O=\sum_k c_k A_k \otimes A_k,
\end{equation}
where $A_k$ are hermitian operators (not necessarily pairwise orthogonal)
and $c_k>0.$ For $O$, ${\rm Tr}(O A \otimes A)\ge 0$ holds for any hermitian
operator $A.$
\begin{itemize}
\item[(i)] If $O$ is symmetric (i.e., it is in the symmetric subspace) and positive semidefinite (e.g., it is an unnormalized
density operator), then it is also PPT.
\item[(ii)] If $O$ is not symmetric, then it is at least permutationally
invariant. If it is also positive semidefinite and of trace one, then it
does not violate the CCNR criterion.
\item[(iii)] If $O$ is not positive semidefinite, then
$O':=O-\Lambda_{\min}(O)\cdot\openone$ is positive semidefinite.
From the previous arguments, it follows that the state $\vr':=O'/{\rm Tr}(O')$
does not violate the CCNR criterion.
\end{itemize}

These ideas can be extended to the multipartite case. Any density matrix
with a bosonic symmetry and of the form
\begin{equation}
\vr_{\rm 4}=\sum_k c_k A_k \otimes A_k\otimes A_k\otimes A_k,
\end{equation}
where $c_k\ge 0,$ is PPT with respect to any $2:2$ (two qubits vs. two qubits) partition. However, it can be non-PPT with respect to the $1:3$ (one qubit vs. three qubits) partitions. Such a density matrix has already been presented by Smolin \cite{S01}
\begin{equation}
\vr_{\rm Smolin}=\frac{1}{16} (\openone^{\otimes 4} + \sigma_x^{\otimes 4}+\sigma_y^{\otimes 4}+\sigma_z^{\otimes 4}),
\end{equation}
where $\sigma_l$ are the Pauli spin matrices. It instructive to consider
the generalized Smolin states defined in
Ref.~\cite{AH05} as
\begin{equation}
\vr_{{\rm Smolin,} n}=\frac{1}{2^{2n}} [\openone^{\otimes 2n} + (-1)^n \sum_{l=x,y,z}\sigma_l^{\otimes 2n}].
\end{equation}
These are permutationally invariant states thus they
do not violate the CCNR criterion with respect to the $n:n$ partition for even $n.$

\section{Entanglement witnesses for symmetric states}

Entanglement witnesses are one of the most important tools for detecting quantum
entanglement \cite{HH09,GT09,H96,H96b,boundwit,boundwitb,boundwitc,boundwitd}. They typically have the form
\begin{equation}
\mathcal{W} =
\sup_{\ket{\psi_1},\ket{\psi_2}} \bra{\psi_1,\psi_2} M \ket{\psi_1,\psi_2} \cdot \openone
- M,\label{witness}
\end{equation}
where $M$ is a hermitian operator. The expectation value of $\mathcal{W}$ is positive on all
separable states, and a negative expectation value signals the presence of entanglement. Without
loosing generality, we can assume that $M$ is positive semidefinite.

For symmetric systems it is known  that all separable states can be decomposed as
\cite{ES02}
\begin{equation}
\vr_{\rm sep,sym}=\sum_k p_k \ketbra{\phi_k} \otimes \ketbra{\phi_k}.
\end{equation}
Thus we can consider a {\it symmetric witness}
\begin{equation}
\mathcal{W}_{\rm sym}
=\sup_{\ket{\psi}} \bra{\psi,\psi} M \ket{\psi,\psi} \cdot\openone - M
\label{witness_sym}
\end{equation}
which simplifies the optimization for the witness. $\mathcal{W}_{\rm sym}$
is in general not a witness for non-symmetric states, but any symmetric
state giving a negative expectation value for $\mathcal{W}_{\rm sym}$
is entangled. Note that this property also holds if we replace the
$\openone$ in Eq.~(\ref{witness_sym}) by the projector onto the
symmetric space $\Pi_S.$

Finally, note that a related way of defining a witness for symmetric states is
\begin{equation}
\mathcal{W}_{\rm sym}'=\sup_{\vr} {\rm Tr}(M\rho \otimes \rho) \cdot\openone - M.
\label{witness_sym2}
\end{equation}
The calculation of $\mathcal{W}_{\rm sym}'$ can  sometimes be numerically easier
than the calculation of $\mathcal{W}_{\rm sym}.$

For the three witnesses we have $\mathcal{W} \le \mathcal{W}_{\rm sym}' \le \mathcal{W}_{\rm sym}.$
It is an interesting question to ask, under which conditions equality holds here. Recently,
it has been shown \cite{robert} that if $M$ is positive semidefinite and symmetric, then $\mathcal{W}=\mathcal{W}_{\rm sym}$
hence all three witnesses are the same.
The same is true if $M$ is a positive permutationally
invariant multi-qubit observable that contains only full correlations terms \cite{robert}.

Next, we will use numerics to obtain entanglement witnesses for a bound entangled
state. There is a theory for witnesses detecting bound entangled states in the literature \cite{boundwit,boundwitb,boundwitc,boundwitd}.
It has also been shown that semidefinite programming
could be used for detecting bound entanglement
and for constructing entanglement witnesses
\cite{numerics,numericsb,numericsc,numericsd,numericse,numericsf}.
We would like to simplify the search for entanglement witnesses for bound entangled states.
We will construct an entanglement witness from
 the Schmidt decomposition of the density matrix of a symmetric bound entangled state.
Semidefinite programming will be used to
show that the witness constructed really detects the state as entangled.

Let us consider a permutationally invariant operator $M.$ It can be written as
\begin{equation}
M=\sum_k c_k M_k \otimes M_k,
\end{equation}
where $M_k$ are local orthogonal observables. Then for obtaining the maximum for
such an operator for states of the form $\vr \otimes \vr$ we have to maximize
\begin{equation}
{\rm Tr}(M \vr \otimes \vr) = \sum_k c_k {\rm Tr} ( M_k \vr )^2.
\end{equation}
Interestingly, one can show that if all $c_k\le 0,$ then this task can be solved straightforwardly by
semidefinite programming \cite{semidefinite}. On the other hand, if all $c_k\ge 0,$ this is not the case,
but the maximum for $\vr \otimes \vr$ is the same as the maximum for $\ket{\psi} \otimes \ket{\psi}.$
This can be seen noting that due to the positive $c_k$'s the target function is convex in $\vr$ and
${\rm Tr}(M \vr \otimes \vr)$ takes its maximum for pure $\vr$'s. The optimization task can be solved
with the method of moments that give a hierarchy of approximations that give better and better lower
bounds on the maximum \cite{moments}.

Based on these ideas now we construct an entanglement witness for the $3\times3$ symmetric
bound entangled state presented in Ref.~\cite{TG09}. For that, we denote the basis states of a
single three-level system by $\ket{\alpha},$ $\ket{\beta}$ and $\ket{\gamma}$ and define the
bipartite symmetric states
$\ket{0}=\ket{\alpha\alpha},$
$\ket{1}=(\ket{\alpha\beta}+\ket{\beta\alpha})/\sqrt{2},$
$\ket{2}=(\ket{\alpha\gamma}+2\ket{\beta\beta}+\ket{\gamma\alpha})/\sqrt{6},$
$\ket{3}=(\ket{\gamma\beta}+\ket{\beta\gamma})/\sqrt{2},$
and
$\ket{4}=\ket{\gamma \gamma}.$ Then, the state 
\begin{align}
\vr&_{3\times 3} = 0.22 \ketbra{0} + 0.176 \ketbra{1} + 0.167 \ketbra{2}\label{rho33}
\\
&+ 0.254 \ketbra{3}
+ 0.183 \ketbra{4} - 0.059  (\ket{3}\!\bra{0}+\ket{0}\!\bra{3})\nonumber 
\end{align}
is bound entangled \footnote{The state Eq.~(\ref{rho33}) was defined originally 
in Ref.~\cite{TG09} in a four-qubit symmetric system,
such that it was bound entangled with respect to the $2:2$ (two qubits vs. two qubits) partition.
It has been discussed that this state can be transformed into a symmetric bound entangled state
of a $3\times 3$ system. Here, we carried out this transformation explicitly
and gave the density matrix for the two-qutrit system.}. To construct the witness, we first obtain a Schmidt decomposition
as
\be
\vr_{3\times 3}=\sum_{k=1}^9 \lambda_k^{(3\times 3)} M_k^{(3\times 3)} \otimes M_k^{(3\times 3)}.
\ee
Here $\lambda_k^{(3\times 3)}$ are ordered in a descending order.
Then, we consider an operator
\be
M_{3\times 3}=\sum_{k=1}^9 f_k M_k^{(3\times 3)} \otimes M_k^{(3\times 3)},
\ee
where $f_k$ are some positive coefficients. With the choice of
$f_k=(\lambda_k^{(3\times 3)})^{\frac{1}{2}}$ for $k=1,2,..,6$ and $f_7=f_8=f_9=0,$ we obtain a witness as
\be
\mathcal{W}_{3\times 3}=0.447775\cdot\openone-M_{3\times 3}.
\ee
Here the constant is determined by looking for the maximum for $M_{3\times 3}$
for states of the form $\vr \otimes \vr,$ with the method of moments using a
third order approximation. The expectation value of the witness $\mathcal{W}_{3\times 3}$
for the state $\vr_{3\times 3}$ is $-0.000753,$ thus it detects the state as entangled.
For the numerical calculations, we used the SeDuMi, YALMIP and QUBIT4MATLAB packages
\cite{sedumi,yalmip,qubit4matlab}.

The fact that $\vr_{3\times 3}$ is detected by the witness $\mathcal{W}_{3\times 3}$ means
that there are no {\it symmetric} separable states with the same values for
$\ex{M_k\otimes M_k}.$ Thus, knowing the expectation values $\ex{M_k \otimes M_k}$ is
enough to detect the state as entangled, provided one knows that the state under consideration
is symmetric. It is an interesting question which part of the symmetric states of two qudits with dimension $d$ can be detected
in this way, by constructing a witness based on the Schmidt decomposition of their
density matrix. Note that there are $d^2$ such two-body correlations, while a symmetric
state can be described by $\sim \frac{d^4}{4}$ real degrees of freedom.

\section{Conclusions}
In this paper, we showed that several entanglement conditions coincide for
symmetric systems. Due to that, a single two-body correlation measurement
is sufficient to detect any non-PPT quantum states in such systems. The
equivalence of the various conditions is connected to the Schmidt decomposition
for symmetric quantum states. The understanding gained this way
can help to construct multipartite bound entangled states. Finally,
we considered constructing entanglement witnesses for symmetric systems.

\acknowledgement

We thank R.~Augusiak, A.~Doherty, P.~Hyllus, T.~Moroder, M.~Navascues, S.~Pironio, R.~Werner
and M.M.~Wolf for fruitful discussions. We thank especially M.~Lewenstein for many useful discussions on bound entanglement. We
thank the support of the EU (OLAQUI, SCALA, QICS), the National
Research Fund of Hungary OTKA (Contract No. T049234),
the Hungarian
Academy of Sciences (J\'anos Bolyai Programme),
the FWF (START
prize) and the Spanish MEC (Ramon y Cajal Programme,
Consolider-Ingenio 2010 project ''QOIT'').

%
%

\end{document}